\documentclass[12pt,preprint]{aastex}

\usepackage{graphicx}

\newcommand{\La}{\mbox{${\rm Ly\alpha}$}}

\newcommand{\Ha}{\mbox{${\rm H\alpha}$}}
\newcommand{\Hb}{\mbox{${\rm H\beta}$}}

\newcommand{\Line}[3]{\Ion{#1}{#2}\,$\lambda$\,#3}
\newcommand{\Lines}[3]{\Ion{#1}{#2}\,$\lambda\lambda$\,#3}
\newcommand{\Ion}[2]{#1{\,\scriptsize #2}}
\newcommand{\Nh}{\mbox{$N_{\rm H}$}}
\newcommand{\Rwd}{\mbox{$R_{\rm wd}$}}

\newcommand{\Acyc}{\mbox{$A_{\rm cyc}$}}
\newcommand{\Lcyc}{\mbox{$L_{\rm cyc}$}}

\newcommand{\Twd}{\mbox{$T_{\rm wd}$}}

\newcommand{\Porb}{\mbox{$P_{\rm orb}$}}

\newcommand{\Msun}{\mbox{$M_{\odot}$}}

\newcommand{\ecsa}{\mbox{$\rm erg\;cm^{-2}s^{-1}\mbox{\AA}^{-1}$}}

\newcommand{\es}{\mbox{$\rm erg\;s^{-1}$}}

\newcommand{\cts}{\mbox{$\rm cts\;s^{-1}$}}

\newcommand{\msy}{\mbox{$\rm \Msun\,yr^{-1}$}}
\newcommand{\kms}{\mbox{$\rm km\,s^{-1}$}}

\begin{document}


\title{Phase-resolved HST/STIS  spectroscopy of the\\ 
exposed white dwarf in the high-field polar AR\,UMa
\altaffilmark{1}}

\author{Boris T. G\"ansicke}
\affil{Universit\"ats-Sternwarte G\"ottingen, Geismarlandstr. 11,
37073 G\"ottingen, Germany}
\email{boris@uni-sw.gwdg.de}

\author{Gary D. Schmidt}
\affil{Steward Observatory, University of
Arizona, Tucson, AZ 85721}
\email{gschmidt@as.arizona.edu}

\author{Stefan Jordan}
\affil{Institut f\"ur Astronomie und Astrophysik, Universit\"at Kiel,
24098 Kiel, Germany}
\email{jordan@astrophysik.uni-kiel.de}

\and 

\author{Paula Szkody}
\affil{Astronomy Department, University of Washington, Seattle, WA
98195}
\email{szkody@astro.washington.edu}

\altaffiltext{1}{Based on observations made with the NASA/ESA Hubble Space
Telescope, obtained at the Space Telescope Science Institute, which is
operated by the Association of Universities for Research in Astronomy,
Inc., under NASA contract NAS 5-26555.}

\begin{abstract}
Phase-resolved {\it HST/STIS\/} ultraviolet spectroscopy of the
high-field polar AR\,UMa confirms that the white dwarf photospheric
\La\ Zeeman features are formed in a magnetic field of
$\sim200$\,MG.
In addition to the
\La\ $\pi$ and $\sigma^+$ components, we detect the forbidden
hydrogen $1s_0\rightarrow2s_0$ transition, which becomes ``enabled''
in the presence of both strong magnetic and electric fields.
Overall, the combined ultraviolet and optical low state spectrum is
similar to that of the single white dwarf PG\,1031+234, in that the
optical continuum has a steeper slope than the ultraviolet continuum
and that the depth of the \La\ Zeeman lines reaches only $30-50$\,\%
of the continuum level. Our attempt in fitting the low state data with
single temperature magnetic white dwarf models remains rather
unsatisfactory, indicating either a shortcoming in the present models
or a new physical process acting in AR\,UMa. As a result, our
estimate of the white dwarf temperature remains somewhat uncertain,
$\Twd=20\,000\pm5000$\,K.
We detect a broad emission bump centered at $\sim1445$\,\AA\ and
present throughout the entire binary orbit, and a second bump near
$\sim1650$\,\AA, which appears only near the inferior conjunction of
the secondary star.  These are suggestive of low harmonic cyclotron
emission produced by low-level ($\dot M\sim10^{-13}$\,\msy) accretion
onto both magnetic poles.  However, there is no evidence in the power
spectrum of light variations for accretion in gas blobs.  The derived
field strengths are $B\sim240$\,MG and $B\ga160$\,MG for the northern
and the southern pole, respectively, broadly consistent with the field
derived from the Zeeman lines.
The observed \La\ emission line shows a strong phase dependence with
maximum flux and redshift near orbital phase $\phi\sim0.3$, strongly
indicating an origin on the trailing hemisphere of the secondary
star. An additional \La\ absorption feature with similar phasing as
the \La\ emission, but a $\sim700$\,\kms\ blueshift could tentatively
be ascribed to absorption of white dwarf emission in a moderately fast
wind.
Finally, the high signal-to-noise {\it STIS\/} data provide important
information on the intergalactic absorption toward AR\,UMa. We derive
a column density of neutral hydrogen of
$\Nh=(1.1\pm1.0)\times10^{18}\rm cm^{-2}$, the lowest of any known
polar, making AR\,UMa an excellent candidate for further EUV
observations.

\end{abstract}

\section{Introduction}
AR\,UMa is a magnetic cataclysmic variable (CV) of the extreme
kind. It was discovered as a very luminous soft X-ray source during
the {\it EINSTEIN\/} slew survey and optically identified as a nearby
($d=88$\,pc), short-period ($\Porb=1.932$\,hr, near the lower edge of
the period gap), and possibly magnetic CV by
\citet{remillardetal94-1}. An analysis of the Harvard and Sonneberg
plate material showed that AR\,UMa spends most of the time in a state
of low accretion activity, $V\sim16.5$, with sporadic high states
reaching up to $V\sim13.5$
\citep{remillardetal94-1,wenzel93-1}. \citet{schmidtetal96-1}
confirmed AR\,UMa as a strongly magnetic CV (polar) with unprecedented
properties: in contrast to all other known polars, AR\,UMa shows
practically {\it no\/} circular polarization during the high accretion
state. Moderate circular polarization is, however, observed during the
low state, and, if interpreted as dichroism in the photosphere of the
white dwarf, indicates a magnetic field $B\ga200$\,MG. Confirmation of
an extremely high field ($B\sim230$\,MG) was provided by the detection
of the \La\ $\sigma^+$ Zeeman component in the {\it IUE\/} low state
spectrum of AR\,UMa \citep{schmidtetal96-1}.  Additional optical
\citep{schmidtetal99-2} and EUV/X-ray observations
\citep{szkodyetal99-1} constrained the binary parameters and
highlighted the central role of the strong magnetic field in the
accretion process.
Summing up, AR\,UMa is the first polar than can compete in magnetic
field strength with the known single high-field white
dwarfs, permitting tests of current theories of magnetically-funneled
accretion onto white dwarfs under extreme conditions. Two additional
systems were recently identified to bridge the magnetic field strength
distribution down to the bulk of the polars; V884\,Her with
$B\approx150$\,MG \citep{schmidtetal01-1} and
RX\,J1007.5$-$2016 with $B\approx90$\,MG \citep{reinschetal99-1}.

We report in this paper the results of high-quality {\it HST/STIS\/}
observations of AR\,UMa which were aimed at an analysis of the
properties of the accreting white dwarf. The new data clearly confirm
a field strength of $\sim200$\,MG, though a detailed modeling of
the spectra fails for yet unknown reasons. The {\it STIS\/} spectra of
AR\,UMa show some evidence for low-level accretion activity during the
low state and such a low interstellar absorption column that the
object is an ideal target for future EUV observations.

\section{\textit{HST/STIS} observations}
{\it STIS\/} observations of AR\,UMa were carried out on 1998 December 9
during five consecutive {\it HST\/} orbits.
The optical monitoring of AR\,UMa reported by \citet{schmidtetal99-2}
shows that the {\it HST\/} data were obtained during a low state
($V\sim16.5$). The last active phase of AR\,UMa prior to the {\it HST\/}
observations was detected in October/November 1998 at an intermediate
magnitude of $V\sim15.5$.
Considering that AR\,UMa was faint in June 1998, this active state
might have lasted for a maximum of $\sim5$ months. Unfortunately,
during July--September, AR\,UMa is too close to the sun,
preventing continuous monitoring.
The bulk of the {\it STIS\/} data were obtained with the G140L
far-ultraviolet (FUV, 1130$-$1710\,\AA) grating; only a single short
G230L near-ultraviolet (NUV, 1660$-$3150\,\AA) exposure was obtained
during the last orbit of the visit (Table\,\ref{t-obslog}).  All data
were taken through the $52\arcsec\times0.2\arcsec$ slit in order to
optimize both throughput and spectral resolution ($R\approx1000$ and
$R\approx500$ for the G140L and the G230L data, respectively).

\placetable{t-obslog}

\subsection{The average \textit{STIS} spectrum}
The strongest features detected in the average {\it STIS\/} spectrum
of AR\,UMa (Fig.\,\ref{f-fuv_average_spec}) are the broad absorption
troughs of the \La\ $\pi$ and $\sigma^+$ Zeeman components. This
confirms the identification of the absorption line observed at
$\sim1300$\,\AA\ in the {\it IUE\/} low state spectrum as
\La\ $\sigma^+$ split in a field of $\sim230$\,MG
\citep{schmidtetal96-1}.  At such field strengths, the $\sigma^-$
component falls shortwards of the wavelength range covered by {\it
STIS\/}, and the strong decrease in flux observed below 1145\,\AA\ is
due to the rapid drop of the {\it STIS\/} sensitivity toward shortest
wavelengths.

The high-quality {\it STIS\/} data ($\mbox{signal-to-noise
ratio}\sim30$ at $1150\,\mbox{\AA}<\lambda<1450\,\mbox{\AA}$ and
$\sim20$ at $1450\,\mbox{\AA}<\lambda<1750\,\mbox{\AA}$ in the average
FUV spectrum) reveals two additional absorption-like structures
centered at $1180$\,\AA\ and $1415$\,\AA, as well as a number of weak,
narrow absorption lines which are most likely of interstellar origin
(Fig.\,\ref{f-fuv_average_spec}). Note that the average spectrum
contains {\it two\/} narrow \La\ absorption lines, centered at
velocities zero and $\sim-700$\,\kms.  Finally, the weakness of the
detected emission lines, \La, \Line{C}{IV}{1550}, and
\Line{Mg}{II}{2800}, confirms that AR\,UMa was in a state of very low
accretion activity at the time of the {\it HST\/} observations.

\placefigure{f-fuv_average_spec}

\subsection{Phase-resolved ultraviolet spectroscopy\label{s-phase_spec}}
The G140L data were obtained over 1.7 binary orbits with strong phase
overlap between the five exposures.  We used the {\tt inttag} and {\tt
calstis} commands of STSDAS to split the G140L datasets
(Table\,\ref{t-obslog}) into 4, 5, 5, 5, and 3 subexposures,
respectively, with individual exposure times ranging from 440\,s to
592.5\,s (corresponding to orbital phase resolutions
$\Delta\phi\approx0.06-0.09$).  The phases of mid-exposure of the 22
spectra were computed using the ephemeris of \citet{schmidtetal99-2},
where $\phi=0$ corresponds to the inferior conjunction of the
secondary star. Finally, the sub-exposures were summed up in five
phase bins to increase the signal-to-noise ratio
(Figure\,\ref{f-fuv_phase_spec}). The spectral features that
noticeably change around the orbit are (1) the $1180$\,\AA\
absorption, being strongest at $\phi\approx0.5$, (2) the \La\
emission, being strongest at $\phi\approx0.3$ and absent shortly
before the inferior conjunction of the secondary star at $\phi\simeq0.9$,
and (3) a very broad ($\sim200$\,\AA) and weak emission centered at
$\sim1650$\,\AA\ that appears between $\phi\simeq0.9-1.1$

\placefigure{f-fuv_phase_spec}

\section{The photospheric white dwarf spectrum}
During a low state, the ultraviolet/optical emission of a polar is
dominated by the photospheric spectra of the white dwarf and of the
secondary star, allowing a direct measurement of quantities such as
the magnetic field strength and the effective temperature of the white
dwarf and the spectral type of the secondary
\citep[e.g.][]{schmidtetal81-1,schwopeetal93-1}.  In AR\,UMa,
\citet{remillardetal94-1} clearly showed that the red end of the
optical spectrum reveals the emission of an M6 dwarf, and derived a
distance of 88\,pc.
For the present analysis, we supplement our {\it HST/STIS\/} spectra
with an optical low-state spectrum obtained on 1998 May 22 with the
CCD Spectropolarimeter \citep{schmidtetal92-2} and 2.3\,m Bok
telescope on Kitt Peak.  The latter data cover the region 4000-8000 A
at a resolution of $\sim12$\,\AA, and result from a 300 s exposure
($\Delta\phi\sim0.04$) centered at $\phi=0.79$.  The use of these
non-simultaneous data is justified by the fact that the low-state
brightness of AR\,UMa has historically shown very little evidence for
change once the system cools from an active phase
\citep{remillardetal94-1,schmidtetal96-1,schmidtetal99-2}, and because
the continuum ($V=16.9$) in the May 1998 data is within 0.1 mag of a
$V$-band measurement obtained 1998 Nov 14, just 3 weeks prior to the
{\it HST\/} observations.

\subsection{White dwarf temperature\label{s-wdtemperature}}

\citet{schmidtetal96-1} fitted the combined {\it IUE\/}/optical low
state spectrum of AR\,UMa using a grid of solar abundance non-magnetic
white dwarf model spectra from \citet{hubeny88-1}. They found that no
satisfying fit can be achieved with a single-temperature model, but
that a 15\,000\,K white dwarf with a 35\,000\,K spot covering 2\,\% of
the white dwarf surface provides a rough match for the data. It is
clear that any analysis based on non-magnetic models is necessarily
prone to large uncertainties, as the strong magnetic field has a major
impact on the bound-free and bound-bound opacities in the white dwarf
atmosphere \citep[e.g.][]{meranietal95-1}.

\placefigure{f-overall}

Figure\,\ref{f-overall} illustrates the discrepancy between the
observed overall spectrum of AR\,UMa and non-magnetic pure hydrogen
white dwarf model spectra. The optical spectrum of AR\,UMa is very
blue, almost Rayleigh-Jeans, and can be described well with a
50\,000\,K non-magnetic white dwarf model spectrum\footnote{The extremely steep
low-state spectrum of AR\,UMa was already noted by
\citet{remillardetal94-1} who compared it in their Fig.\,1 with the
low-state spectrum of AM\,Her, which has $\Twd\approx20\,000$\,K
\citep{heise+verbunt88-1,gaensickeetal95-1}.}.  In contrast to this,
the {\it STIS\/} FUV/NUV spectrum is relatively flat, suggesting
$\Twd\sim15\,000$\,K.
If we assume a distance of 88\,pc \citep{remillardetal94-1}, fitting
the flux of the ultraviolet data with the 15\,000\,K model yields a
white dwarf radius of $\Rwd=10^9$\,cm, which corresponds to emission
from the entire white dwarf. In contrast, the 50\,000\,K fit to the
optical data can only be interpreted as emission from a small spot
on the white dwarf.
This mismatch of the ultraviolet and the optical continuum is actually
{\it the opposite\/} of what is expected from a white dwarf plus hot
spot model. In such a configuration, the ultraviolet range is
dominated by the hot spot and has a steeper slope than the optical 
range which reveals the cooler underlying white dwarf
\citep[e.g.][]{gaensickeetal00-1}.

Very similar problems were encountered by \citet{schmidtetal86-2} when
interpreting the ultraviolet and optical spectrum of the {\it
single\/} magnetic white dwarf PG\,1031+234 ($B\sim500$\,MG). Also in
this star, which certainly has no accretion-heated pole cap, the
ultraviolet continuum is relatively flat, corresponding to
$\Twd\approx15\,000$\,K, while the optical continuum is steeper,
indicating $\Twd\approx25\,000$\,K.

We attempted a better description of the observed ultraviolet plus
optical low state spectrum of AR\,UMa using the magnetic white dwarf
model spectra of \citet{jordan92-1} \citep[see
also][]{putney+jordan95-1,burleighetal99-1},  assuming a magnetic
field of $B=200$\,MG \citep{schmidtetal96-1}.  We computed the spectra
for an angle of $20^{\circ}$ between the line-of-sight and the
magnetic axis, which corresponds to the closest approach of the
magnetic pole to the observer \citep{schmidtetal99-2}. The positions
of the \La\ components are well reproduced, confirming the field
strength derived by \citet{schmidtetal96-1}.
Figure\,\ref{f-overall} shows the low state data of AR\,UMa along with
magnetic white dwarf model spectra for $\Twd=15\,000$, 20\,000, and
25\,000\,K. The mismatch between the optical and the ultraviolet
continuum is somewhat alleviated compared to the fit with non-magnetic
models, but, again, no single-temperature model provides a
satisfactory fit to the overall spectrum of AR\,UMa. The scaling
applied to the magnetic model spectra in Fig.\,\ref{f-overall} implies
white dwarf radii of $5.1\times10^8$\,cm and $4.3\times10^8$\,cm for
the 20\,000\,K and the 25\,000\,K models, respectively, for a distance
of 88\,pc \citep{remillardetal94-1}.  These values are compatible with
the assumption that the low state spectrum of AR\,UMa is dominated by
emission from the magnetic white dwarf.

\placefigure{f-fuv_mwd}

We also attempted to model the {\it STIS\/} FUV data alone with our
magnetic white dwarf spectra, again with only limited success:
Fig.\,\ref{f-fuv_mwd} shows the best match with $\Twd=17\,000$\,K and
$\Rwd=8.2\times10^8$\,cm (for $d=88$\,pc). While the FUV continuum is
well described, the model over-predicts the optical flux in the $V$
band by a factor of $\sim2$.
Interestingly, the Zeeman lines of the model spectra are more deeply
modulated and have, hence, larger equivalent widths than the observed
lines. Again, this is very similar to the observations of
PG\,1031+234, where the \La\ $\sigma^+$ component is modulated only to
$\sim$50\,\% of the continuum flux. \citet{schmidtetal86-2} argued
that the Zeeman transitions can only absorb one of two orthogonal
polarization modes of the outgoing flux and, hence, the depth of a
saturated line will reach maximally 50\,\% of the continuum level.

A quantitative prediction of the line depth and strength is
complicated, as magneto-optical effects (Faraday rotation and Voigt
effect) change the polarization properties of the absorption
coefficients throughout the atmosphere. In addition, there is at
present no reliable theory for the Stark broadening of the individual
Zeeman components in the presence of a strong magnetic field. We
carried out two exploratory model atmosphere calculations for
$\Twd=20\,000$\,K, $B=200$\,MG, once including and once excluding the
magneto-optical effects. The strongest transition, the \La\ $\pi$
component, is modulated to $\sim50$\,\% in the spectrum that was
computed without magneto-optical effects, just as expected according
to \citet{schmidtetal86-2}. However, the $\pi$ component is almost
100\,\% modulated in the spectrum that was computed including the
magneto-optical effects. Similar variations are observed in the
$\sigma^+$ and $\sigma^-$ components. We have, thus, to conclude that
at present it is not possible to self-consistently model the spectrum
of the magnetic white dwarf in AR\,UMa (or that of PG\,1031+234).

Just for completeness, we note that AR\,UMa as an interacting binary
offers two other possibilities to explain the low depth of
the \La\ lines: either another component in the binary contributes
significantly to the FUV flux, or the white dwarf atmosphere is heated
by accretion, resulting in a flatter vertical temperature gradient and
weaker absorption lines (as observed e.g. in AM\,Her during a high
state; \citealt{gaensickeetal98-2}). However, neither
hypothesis is very appealing, as AR\,UMa was observed with {\it
HST\/} during a state of very low accretion activity.

\subsection{\label{s-1180}
The 1180\,\AA\ absorption line: a forbidden \La\
$\mathbf{1s_0\rightarrow2s_0}$ component in the atmosphere of the white dwarf?}
In addition to the \La\ $\pi$ and $\sigma^+$ components, a relatively
broad (FWHM$\sim$11\AA) absorption feature centered at
$\approx1180$\,\AA\ appears during $\phi=0.3-0.7$ without noticeable
variation in wavelength (Fig.\,\ref{f-fuv_phase_spec}).  Comparison
with the phase-resolved {\it HST/GHRS\/} spectra of the single
magnetic white dwarf RE\,J0317--853 suggests that the 1180\,\AA\
absorption is due to the normally forbidden dipole transition of
hydrogen $1s_0\rightarrow2s_0$, which becomes increasingly more
probable in the presence of both strong magnetic and electric fields,
a situation encountered in the high-density atmospheres of magnetic
white dwarfs \citep{burleighetal99-1}. In RE\,J0317--853, the
$1s_0\rightarrow2s_0$ absorption is attributed to the weaker pole with
$B\sim200$\,MG. In AR\,UMa, the 1180\,\AA\ is strongest at
$\phi=0.3-0.5$ when the northern magnetic pole $B\sim240$ is best
visible (see Sect.\,\ref{s-cyc_model}).

Both the wavelength and the oscillator strength of the $1s_0
\rightarrow 2s_0$ component depend on the choice of the electric
field. The oscillator strength of this ``forbidden'' feature increases
with the electric field at the expense of the normal $\pi$ ($1s_0
\rightarrow 2p_0$) component. Since we have at present the necessary
atomic data only for a limited range of electric field strengths we
assumed in our model spectrum calculation (Fig.\,\ref{f-fuv_mwd}) the
same electric field of $10^8 $ V/m as for RE\,0317--853
\citep{burleighetal99-1}. 
A detailed modeling of the $1s_0 \rightarrow 2s_0$ feature, especially
of its phase-dependence, will be possible only once the necessary
atomic data for a large range of fields are available.

The observed $1s_0 \rightarrow 2s_0$ in AR\,UMa is stronger than
in the model, in fact of similar strength as the observed $\pi$ ($1s_0
\rightarrow 2p_0$) component! Therefore, the electric field in the
atmosphere of AR\,UMa should be even higher than in that of
RE0317--853, which could also explain the slight difference in
wavelength between model and observation.
Unfortunately, we have no good physical explanation for the origin of
such a strong electric field in the atmosphere of AR\,UMa: The
strength of the electric micro field expected from the disturbing
particles, ions and electrons, scales with the electron density as
$N_e^{2/3}$, which implies that the electric field increases with the
effective temperature because of a higher degree of ionization. From
this simple assumption, the electric field strength in AR\,UMa should
be a factor of $\sim3$ {\it lower} than in RE\,0317--853. A similar
conclusion holds for the thermally induced motional Stark effect or
the generation of electric fields from a rotating dynamo.

Finally, we add as a cautionary remark that, as the white dwarf in AR\,UMa is
accreting metal-rich material from its secondary, we cannot rule out
that heavy elements could contribute to the observed line
spectrum. \citet{schmidtetal96-1} already argued that a number of
absorption lines in the optical spectrum could be from elements
other then hydrogen. In fact, a broad (FWHM$\sim$8\,\AA) absorption is
observed in the {\it STIS\/} spectrum at $\sim1335$\,\AA, as well as a
few narrow features that cannot be explained with interstellar
absorption (see Sect.\,\ref{s-ism}), e.g. 1227.7\,\AA\ and
1325.6\,\AA. We note that the 1180/1335\,\AA\ features coincide with
\Line{C}{III}{1176} and \Line{C}{II}{1335}, which are very strong in
$\sim20\,000-30\,000$\,K metal-rich atmospheres. However, in the
absence of theoretical descriptions of metal transitions in strong
magnetic fields and as no other typically strong absorption, e.g. near
\Line{Si}{II}{1260,65} or \Line{Si}{III}{1300}, is observed, we
consider the hydrogen $1s_0\rightarrow2s_0$ transition to be the most
likely explanation of the $1180$\,\AA\ feature.

\placefigure{f-cyc_model}

\subsection{The broad 1415\,\AA\ absorption}
The average spectrum (Fig.\,\ref{f-fuv_average_spec}) contains a
strong and broad ($\sim70$\,\AA) absorption-like feature centered at
$\sim1415$\,\AA. From Fig.\,\ref{f-fuv_phase_spec} it appears that
this absorption does not significantly depend on the orbital
phase. We envision several possible mechanisms that could be the
source of this feature.

\subsubsection{Quasi-molecular $\mathrm{H}_2^+$ absorption.} 
In non-magnetic DA white dwarfs with effective temperatures below
$\sim20\,000$\,K, absorption by quasi-molecular hydrogen causes a
broad line centered at $\sim1400$\,\AA\ \citep{koesteretal85-1}.
\citet{schmidtetal96-1} fitted the combined {\it IUE\/}/optical low state
spectrum of AR\,UMa with a 15\,000\,K white dwarf with a small
35\,000\,K spot (see also the discussion in
Sect.\,\ref{s-wdtemperature}). Such a low temperature of the white
dwarf makes the $\mathrm{H}_2^+$ hypothesis appear viable. However,
present theory can not predict the position/shape of the
$\mathrm{H}_2^+$ absorption in a strong magnetic field. Also on
observational grounds, the evidence for $\mathrm{H}_2^+$ absorption in
magnetic white dwarfs is meagre: \citet{gaensickeetal00-1} detected
${H}_2^+$ absorption in only two out of 12 {\it single\/} magnetic white
dwarfs that were observed with {\it IUE\/}, and in none of the accreting
magnetic white dwarfs in polars observed so far in the ultraviolet.

\subsubsection{\label{s-zeeman_balmer}Zeeman absorption by Balmer lines}
In extremely strong magnetic fields several transitions of the Balmer
line series are shifted by up to thousands of \AA\ into the
ultraviolet \citep[e.g.][]{henry+oconnell85-1}. 
It may, hence, be also possible that the observed absorption near
1415\,\AA, the phase-dependent flux variation around 1650\,\AA, and
other structures in the long-wavelength portion of the {\it STIS\/}
spectrum, are related to Balmer absorption.  Indeed, our model spectra
(Fig.\,\ref{f-overall}) qualitatively reproduce the undulations
observed in the NUV spectrum of AR\,UMa, however they do not contain
significant structure in the FUV range (apart from \La).

\subsubsection{Cyclotron emission\label{s-cyc_model} -- ongoing accretion} 
Another possible interpretation arises if the structure near
$1415$\,\AA\ is not due to absorption, but if there is a rather
broad {\it emission\/} bump redward of it, peaking at
$\sim1445$\,\AA. In the context of accreting magnetic white dwarfs,
cyclotron emission is an immediate candidate for the origin of such a
broad emission line. As the shape of the feature does not change
significantly around the orbit we would have to assume that it originates in
a hot accretion plasma near the northern magnetic pole, which remains
constantly in view \citep{schmidtetal99-2,szkodyetal99-1}.
The cyclotron hypothesis finds some support in the fact that an
additional bump centered at $\sim1650$\,\AA\ appears in the FUV
spectrum near $\phi\approx0$. The spectral shape of this bump is most
clearly seen in the flux difference $(\phi=0.5)-(\phi=0.0)$, see
Fig.\,\ref{f-cyc_model}, right panel, bottom curve. The appearance of
the 1650\,\AA\ bump coincides with the phase when the southern
accretion region rotates into view
\citep{schmidtetal99-2,szkodyetal99-1}, and might be related to
accretion onto the southern magnetic pole. We recall here that the
southern pole is the more active one during the high state.

We model the hypothetical cyclotron features with the emission of an
isothermal plasma slab in a strong magnetic field. The intensity
spectrum is simply given as
$I_{\lambda}=B_{\lambda}(1-e^{-\tau_{\lambda}})$, with $B_{\lambda}$
the Planck function and $\tau_{\lambda}$ the wavelength-dependent
optical depth.  We follow the general approach \citep[see
e.g.][]{wickramasinghe88-1} and express
$\tau_{\lambda}=\Lambda\varphi_{\lambda}$ with the $\Lambda$ the
dimensionless size parameter and $\varphi_{\lambda}$ the dimensionless
absorption coefficient \citep[computed according
to][]{chanmugam+dulk81-1,thompson+cawthorne87-1}. Free parameters of
this model are the angle $\vartheta$ between the line of sight and the
magnetic field line threading the emitting plasma, the magnetic field
strength $B$, the plasma temperature $kT$, and the size parameter
$\Lambda$.
The detection of a single cyclotron harmonic does not allow an
independent estimate of the magnetic field strength. We computed,
therefore, the cyclotron model spectra for a range of field strengths
that are compatible with $B\simeq200$\,MG, as derived from the
Zeeman-split \La\ line (Sect.\,\ref{s-wdtemperature}).

In a first step, we modeled the 1445\,\AA\ emission bump in the FUV
spectrum obtained at $\phi=0.5$, as it is least contaminated by the
emission bump at 1650\,\AA. We obtain a satisfactory fit for
$B=240$\,MG, $kT=0.45$\,keV, $\vartheta=35^{\circ}$, and
$\log(\Lambda)=7.0$ (bottom curve of Fig.\,\ref{f-cyc_model}, left
panel). For these parameters, the 1445\,\AA\ bump corresponds to the
third harmonic of the cyclotron emission. The model spectrum predicts
some emission in the second harmonic at $\sim2230$\,\AA, and very
little flux (a few $10^{-17}\ecsa$) in the fundamental frequency at
$\sim4470$\,\AA.  The NUV spectrum shows a small change in slope at
$\sim2100$\,\AA, which might be related to flux from the second
harmonic.  Subtracting the cyclotron model from the observed FUV
spectrum results in an almost smooth continuum
(Fig.\,\ref{f-cyc_model}, left panel). The remaining structure is not
too surprising, as the accretion region will be characterized by a
temperature and density structure which can not fully be described by our
simple model.

Subsequently, we modeled the 1650\,\AA\ emission bump which appears at
$\phi\approx0.0$ with a second cyclotron spectrum for $B=160$\,MG,
$kT=0.45$\,keV, $\vartheta=35^{\circ}$, and $\log(\Lambda)=9.9$
(Fig.\,\ref{f-cyc_model}, bottom panel). This time, the FUV cyclotron
bump corresponds to the fourth harmonic. Some emission is predicted in
the third ($\sim2230$\,\AA) and second ($\sim3350$\,\AA) harmonics,
and practically no power is expected in the fundamental frequency
($\sim6700$\,\AA).

For both emission regions, the derived model parameters imply an
emitting area $\Acyc\sim10^{14}\mathrm{cm}^2$ and a
luminosity $\Lcyc\sim10^{30}\es$. For a canonical white dwarf
mass and radius (0.6\,\Msun, $8.4\times10^8$\,cm), the cyclotron
luminosity corresponds to an accretion rate of
$\sim10^{13}\mathrm{g\,s^{-1}}$, or $\sim1.7\times10^{-13}$\,\msy.

In summary, if the 1445\,\AA/1650\,\AA\ bumps are really due to
cyclotron emission, the implication is that a very low inflow of
material feeds both the northern and the southern pole during the low
state.
The low cyclotron flux that both poles emit in the fundamental
harmonics is consistent with the non-detection of low-state cyclotron
emission (both in polarization and in flux) in the optical range
\citep{schmidtetal96-1}.
The derived cyclotron luminosity is somewhat larger than the limit on
the low state hard X-ray luminosity obtained from ASCA data
\citep{szkodyetal99-1}. The plasma temperature that we derive is
extremely low, but compatible with the combination of a low accretion
rate and the efficient cyclotron cooling of the post shock flow in a
strong field \citep[e.g.][]{beuermann+woelk96-1}.
\citet{fischer+beuermann00-1} present updated model temperature
structures for accretion columns in polars and predict indeed
low-temperature ($kT\la1$\,keV) bremsstrahlung from the accretion
region for such a combination.

The field strengths derived from the cyclotron emission may be
considered an indication that the field geometry is not that of a
simple dipole. Such conclusions are not unusual in the studies of
single and accreting magnetic white dwarfs
\citep[e.g.][]{wickramasinghe+martin79-1,burleighetal99-1,schwopeetal93-1}.
However, the derived 160\,MG should be considered as a lower limit on
the field strength of the southern pole, as the magnetic south pole
remains eclipsed by the white dwarf throughout the entire binary orbit
\citep{schmidtetal99-2}.
The detection of cyclotron radiation from the southern accretion
region implies that this region is rather extended -- which is not too
surprising: because of the high magnetic field strength in AR\,UMa
material is already magnetically threaded  near the secondary star and
may, consequently, be distributed over large areas near the magnetic
poles of the white dwarf.
The observed weak orbital dependence of the 1445\,\AA\ bump requires
that the northern cyclotron emitting region must be located very close
($5^{\circ}-10^{\circ}$) to the rotation axis of the white dwarf to
minimize the orbital variation of $\vartheta$.
\citet{schmidtetal99-2} derived from a number of
observational constraints 
a colatitude of the magnetic pole $10^{\circ}\la \delta_\mathrm{B} \la
35^{\circ}$. Keeping in mind that the actual accretion region(s) in
polars are usually offset by some degree from the magnetic pole(s),
these conclusions are consistent.

\section{Narrow absorption lines of low ionization species: ISM\label{s-ism}}
Besides the broad \La\ Zeeman components, the mean spectrum of AR\,UMa
displays a number of narrow absorption lines of various strengths.
An enlargement of the FUV average spectrum
(Fig.\,\ref{f-fuv_average_spec}) shows that some of the more
pronounced narrow absorption features coincide with the strongest
interstellar absorption lines that are expected in the spectrum of a
nearby galactic ultraviolet-bright source
\citep[e.g.][]{gaensickeetal98-2,holbergetal99-1,maucheetal88-1}.  The
most convincing detections are \La, \Line{Si}{II}{1260.4}, and
\Line{C}{II}{1334.5} (the latter being embedded in a much broader
--~FWHM$\sim$8\,\AA~-- absorption trough that can not be of
interstellar origin, see Sect.\,\ref{s-1180}). The interstellar origin
of these narrow features is supported by the fact that only the blue
component of the \Lines{Si}{II}{1260.4, 1264.7} doublet is present in
the spectrum of AR\,UMa. The red component corresponds to a transition
from an excited level which is not populated in the interstellar
medium.
The interstellar lines of \Line{Si}{II}{1260.4} and
\Line{C}{II}{1334.5} have equivalent widths of $30-50$\,m\AA, about
half of what is  observed in AM\,Her \citep{gaensickeetal98-2}.
For completeness, we show in Fig.\,\ref{f-fuv_average_spec} the
positions of other interstellar transitions of somewhat lower strength
than the observed \Ion{Si}{II} and \Ion{C}{II} lines,
i.e. \Lines{Si}{II}{1190.70,1193.29}, \Line{N}{I}{1200},
\Line{O}{I}{1302.2}, and \Line{Si}{II}{1304.4}.
We can not claim a significant detection of any of these lines, but we
note that all transitions coincide with absorption dips at the noise
level.

The observed interstellar \La\ absorption can be used to derive an
estimate of the column density of neutral hydrogen along the line of
sight to AR\,UMa. 
We use for this purpose the FUV spectrum obtained shortly before
inferior conjunction of the secondary ($\phi=0.9$), where the
interstellar \La\ profile is least contaminated by the phase-dependent
\La\ emission (Fig.\,\ref{f-fuv_phase_spec}).
We fitted a pure damping profile \citep{bohlin75-1} folded with a
1.2\,\AA\ FWHM Gaussian to the observed \La\ absorption line
(Fig.\,\ref{f-lyalpha}, top panel). The resulting column density of
{\em neutral\/} hydrogen is $\Nh=(1.1\pm1.0)\times10^{18}\rm
cm^{-2}$. This value is even lower than that derived by
\citet{szkodyetal99-1} from EUVE observations,
$\Nh=(6-10)\times10^{18}\rm\,cm^{-2}$.
The higher column density determined from X-ray data indicates the
presence of material along the line of sight, presumably within the
binary system, in which hydrogen is ionized to a high degree while the
other elements are only partially ionized and still contribute to the
soft X-ray absorption. We note in passing that AR\,UMa is the polar
with the {\it lowest\/\/} column density identified so far. As it is also
a very bright EUV source, it is {\it the\/} candidate for future
EUV/soft X-ray observations, e.g. searching for spectroscopic evidence of
metals in the footpoint of the accretion column.

\placefigure{f-lyalpha}

\section{Evidence for a wind in AR\,UMa?\label{s-wind}}
The FUV spectrum of AR\,UMa (Fig.\,\ref{f-fuv_average_spec}) displays
phase-dependent \La\ emission, as well as a weak absorption centered
at $\sim1213$\,\AA\ (in addition to the interstellar \La\ absorption).
Figure\,\ref{f-lyalpha}, bottom panel, shows the \La\ region of the
FUV spectra, corrected for an interstellar absorption of
$\Nh=1.1\times10^{18}\mathrm{cm^{-2}}$.

We fitted the \La\ emission in all 22 short subexposures
(Sect.\,\ref{s-phase_spec}) with Gaussians in order to derive the
variation of the flux and the velocity of the line
(Fig.\,\ref{f-lyalpha_flux}). The parameters derived for the spectra
covering $\phi=0.5-1.0$ are very uncertain, as the \La\ emission
almost vanishes during these phases. A sine fit to the emission line
velocities results in a half-amplitude $K=\sim367\pm58$\,\kms, a mean
velocity $\gamma=222\pm43$\,\kms, and maximum redshift occurring at
$\phi=0.31\pm0.03$. These parameters are, apart from the higher
$\gamma$ velocity in the ultraviolet data, very similar to those
derived from the radial velocities of \Ha\ and \Hb\ observed in 
the low state.
Every case of strong \La\ emission occurs with a significant redshift,
suggesting that the emitting region is moving away from us at
$\phi\approx0.3$. The only component in the binary which is doing this
at the given orbital phase is the secondary star (a weak stream would
still be expected to show a blueshift or no net velocity near this
phase). It seems, therefore, plausible to identify the secondary star
as the origin of the \La\ emission. The width of the \La\
emission ($\mathrm{FWHM}\approx570$\,\kms) is, however, significantly
larger than the rotational broadening on the tidally locked secondary
star, indicating that the emission region is not stationary on or near 
the surface of the secondary.

\placefigure{f-lyalpha_flux}

The phase of maximum \La\ flux is $\phi\approx0.3$, which corresponds
to the best view onto the trailing hemisphere of the irradiated
secondary star. The EUVE spectrum of AR\,UMa shows strong
\Line{He}{II}{304} emission at $\phi\approx0.4$, indicating that the
secondary star is, at least during the high state, not irradiated
uniformly, i.e. the leading hemisphere is probably shielded by the
accretion stream \citep{szkodyetal99-1}.
The \La\ emission appears to become increasingly double peaked between
$\phi\approx0.3-0.5$, with the ``dip'' in the line core centered at
$1216.5$\,\AA. This could indicate either additional absorbing neutral
hydrogen in the line of sight with a velocity of $\sim250$\,\kms, or
that the \La\ emission consists of two components, which would
also be an explanation for the relatively large width of the \La\
emission (see above).

The absorption component at $\sim1213$\,\AA\ is also strongest at
$\phi\approx0.3$, favoring a common origin of both, the emission and
the absorption feature. Identifying the absorption with \La\ yields a
velocity of $\sim-700$\,\kms. The width of the absorption feature is,
in contrast to the \La\ emission, relatively narrow and indicates a
low velocity gradient in the absorbing material. In addition, the
wavelength of the absorption component does not vary significantly
with the orbital phase.  The general shape of the \La\
absorption/emission compound is reminiscent of a P\,Cygni profile,
and, hence, very suggestive of the existence of a moderately
fast wind. In a somewhat speculative way, we attribute this \La\
feature to a wind originating from the asymmetrically irradiated
secondary star. The narrow, almost stationary absorption corresponds
in this case {\it not\/} to the absorption of light from wind source
-- the secondary star (as it does not contribute at all to the
ultraviolet flux), but could be related to absorption of light from
the white dwarf by the intervening wind material.

One may be prone to suggest the accretion stream as an alternative
origin of the \La\ absorption/emission compound. This possibility
seems, however, rather unlikely: around $\phi\approx0.3$, only the
footpoint of the accretion stream feeding the northern pole could
absorb light from the white dwarf surface, but has at this point a
{\it receding} velocity of several 1000\,\kms.

\placefigure{f-lightcurves}

\section{Ultraviolet light curves}
All the {\it STIS\/} data (Table\,\ref{t-obslog}) were obtained using
the MAMA detectors in the time-tagged mode. The prime data product of
the detectors are photon event tables which list the arrival time $t$
and the detector coordinates ($x, y$) for each registered photon. The
time resolution of the MAMA's is $125\mu\,\rm s$. This data format
permits the extraction of light curves in arbitrary wavelength bands
and with any desired time resolution.

\subsection{Orbital flux modulation}
We chose to produce light curves from the G140L observations in the
four wavelengths listed in Table\,\ref{t-lc_par}. The bands were selected
to avoid the geocoronal \La\ emission\footnote{which is strong in the
raw data, but successfully subtracted in the calibrated spectrum
produced by the STSDAS calibration pipeline} and to provide roughly
even sampling over the observed FUV range.

\placetable{t-lc_par}

The first step in extracting the light curves is to select
appropriate ($x, y$) regions in the raw two-dimensional detector image
which contain the desired wavelength range of the object spectrum, as
well as empty background regions above and below the spectrum. We used
boxes 35 pixels wide in the cross-dispersion direction to extract the
source photons.  Care was taken to exclude the area of the detector
shadowed by the repeller wire when selecting the background regions.
In a second step, two new event tables were created for each wavelength
band from all the photon arrival times included in the previously
defined source and background ($x, y$) region(s),
respectively. Finally, these two event tables were sampled in equally
spaced time bins, i.e. light curves, and the background light curves
were subtracted from the source light curves, scaled appropriately for
the detector areas used in the extraction process. The light curves
were converted from \cts\ to fluxes by scaling to the average flux in
the selected wavelength band.
Figure\,\ref{f-lightcurves} shows the background subtracted
phase-folded light curves in the wavelength bands a--d, sampled in
120\,s.

In order to derive amplitudes and phases of the modulations we fitted
simple sine functions of the form $A\,\sin[(\phi-\phi_0)2\pi]+O$ to
the FUV light curves, with $A$ the half-amplitude of the sine wave,
$\phi_0$ the phase offset, and $O$ the mean flux. The best-fit
parameters are reported in Table\,\ref{t-lc_par}. 

At first glance, the sinusoidal FUV modulation in AR\,UMa is
reminiscent of the FUV light curves observed in a number of polars
during both high and low states and interpreted by rather large
``warm'' spots near the accretion regions on the white dwarfs
\citep[e.g.][]{demartinoetal98-1,gaensickeetal98-2,stockmanetal94-1}.
In AR\,UMa, the phases of maximum flux
($\phi_\mathrm{max}=\phi_0+0.25$) vary between 0.95--1.06, which is
only slightly later than the phase of maximum EUV flux during the high
state. \citet{szkodyetal99-1} convincingly argued that the EUV flux
maximum is due to the appearance of the main accreting (southern) pole
during the phase interval $\sim0.7-0.1$.

It seems, hence, very tempting to attribute the observed FUV
modulation to a hot spot near the southern accretion region, which is
either deeply heated from the previous high state \citep[the cooling
time scale of the accretion regions is
$\sim$months,][]{schmidtetal96-1} or heated by ongoing low level
accretion (Sect.\,\ref{s-cyc_model}).  However, two points argue
against this interpretation:

(1) a spot near the southern magnetic pole would be eclipsed by the
body of the white dwarf for a large part of the orbital phase,
resulting in a flat-bottomed light curve (similar to the EUVE light
curve, \citealt{szkodyetal99-1}), which contrasts with the observed
sinusoidal shape of the FUV modulation. To match the observed shape of
the FUV light curve, a southern spot would have to be ridiculously
large, covering $\sim70-90\%$ of the white dwarf, in which case it
would be more appropriate to speak of a {\it cold\/} northern polar
cap. Thus, if due to a temperature variation over the white dwarf
surface, the observed FUV modulation would have to be ascribed to a
hot polar cap on the northern pole. But a northern pole cap
cannot reproduce the observed modulation, as its aspect changes only
little with the orbital phase.

\placefigure{f-modulation_amplitude}

(2) the amplitude of the flux modulation is apparently not a simple
function of wavelength, but shows a minimum in the
1323.2--1520.3\,\AA\ band (Table\,\ref{t-lc_par}).  This is in strong
contrast to the FUV modulation observed in other polars, where the
amplitude of the FUV modulation has always been found to monotonically
increase toward shorter wavelengths.  To investigate the wavelength
dependence of the FUV modulation in AR\,UMa, we extracted from the
time-tagged photon lists a total of 18 phase-folded light each one
covering $\sim30$\,\AA. The signal-to-noise ratio of these light
curves is of course lower than that of the four broad-band light
curves, and we computed the relative modulation of a given light curve
as the ratio of the standard deviation of count rate to the mean count
rate in the corresponding wavelength band
(Fig.\,\ref{f-modulation_amplitude}).

As already suggested by the broad-band light curves described above, a
strong modulation is observed only for $\lambda<1300$\,\AA\ and
$\lambda>1600$\,\AA. The maximum modulation ($\sim10$\%) coincides
with the $\pi$ component of \La, the minimum with the ``1415\,\AA''
absorption feature.  Thus, cyclotron emission from the northern pole
may dilute the modulation somewhat
(Sect.\,\ref{s-cyc_model}). However, the cyclotron flux predicted from
our model alone is too low to explain the observed decrease of the FUV
modulation by a factor $\sim4$.

Summing up, the interpretation of the FUV modulation observed with
{\it STIS\/} cannot follow in a straightforward way the results
obtained for other polars. Even though a hot spot may be
present in AR\,UMa, and may contribute to the observed flux
modulation, it appears likely that different phenomena cause the
observed modulation at the blue and the red ends of the {\it STIS\/}
bandpass. 
At short wavelengths the strong modulation coincides with the
\La\ Zeeman absorption structure which sensitively depends on the
field dependent photospheric opacities. A slight orbital variation of
the projected magnetic field may thus result in a variation of the
flux in this band.
At long wavelengths, the broad emission bump
which appears at $\phi=0.9-1.1$ and which we interpreted as cyclotron
emission best explains the observed flux modulation.

\subsection{Probing for blobby accretion}
In Sect.\,\ref{s-cyc_model} we suggested that the bumps observed in
the FUV spectrum of AR\,UMa are due to cyclotron emission, implying
that the white dwarf is accreting at a tiny rate even during the low
state. We used the {\it STIS} time-tagged photon stream to probe for
evidence of ``blobby accretion'', i.e. for stochastic variability 
in excess to pure Poisson noise. 
For this analysis, we extracted a count rate light curve from the {\it STIS}
photon stream in an analogous fashion as described above, using the
entire G140L wavelength range except for a narrow range around the
geocoronal \La\ emission. The background subtracted light curve,
sampled in 10s bins, shows a sinusoidal modulation with an
half-amplitude of 3.6\,\%. We computed the discrete power spectrum of
this light curve using the MIDAS Time Series Analysis context
(Fig.\,\ref{f-power}). Clearly present is the orbital period plus a
number of aliases caused by the uneven sampling of the orbital flux
modulation.
In order to probe for additional non-Poissonian power, we composed
synthetic time-tagged data sets with the same mean count rate and the
same coverage as the STIS data. The synthetic count rates were
modulated by a sine wave with the parameters derived from the best
sine-fit to the STIS count rate light curve. The synthetic data sets
were sampled in 10s bins and, finally, we computed discrete power
spectra from the synthetic count rate light curves. The power spectra
for the synthetic data sets reproduce very well the power spectrum
obtained from the real data, both the various spikes corresponding to
aliases of the orbital period and the noise level at the high
frequency end frequencies (Fig.\,\ref{f-power}). From this comparison,
we conclude that the ultraviolet data provide no significant evidence
for short-term fluctuations of the count rate due to individual
accretion events down to a count rate of $\sim0.4$\,\cts,
corresponding to 0.1\,\% of the mean count rate.

\section{Summary}
Our analysis of the {\it HST/STIS\/} observations of AR\,UMa lead to
the following results in our understanding of this unique accretion
physics and plasma laboratory.

\begin{enumerate}
\item We clearly confirm the earlier {\it IUE\/} detection of the
photospheric \La\ $\sigma^+$ absorption, and with an implied magnetic
field strength of $\sim200$\,MG AR\,UMa may well be called ``the king
of the polars''.
Alas, our state-of-the-art magnetic white dwarf model spectra fail to provide 
a satisfactory description of either the detailed \La\ Zeeman profiles or 
the ultraviolet/optical spectral energy distribution, resulting in only a 
rough temperature estimate, $\Twd=20\,000\pm5000$\,K.

\item The uncooperative (pole-on) viewing geometry prevents a detailed
mapping of either the magnetic field topology or a potential
temperature variation over the white dwarf surface. Additional
information can probably be obtained from the phase-dependency of the
``forbidden'' hydrogen $1s_0\rightarrow2s_0$ transition.
However, detailed modeling must await the necessary atomic data.

\item As a consequence of the high field strength, the fundamental 
cyclotron frequency falls in the optical wavelength band.  However, 
there is some evidence in the ultraviolet for low-harmonic cyclotron emission 
originating from near the magnetic pole and corresponding to a low-state 
accretion rate of $10^{-13}$\,\msy.
Remnant activity in the system during the low state is also indicated
by the presence of \La\ emission.  The observed orbital variation of
flux and velocity leads us to attribute this emission to the secondary
star, possibly to a wind emanating from its trailing hemisphere.

\item The brightness and extremely low interstellar column mark
AR\,UMa as the best polar for future FUV/EUV observations.

\end{enumerate}

\acknowledgements{Support was provided through NASA grant GO-7397 from
the Space Telescope Science Institute, which is operated by the
Association of Universities for Research in Astronomy, Inc., under
NASA contract NAS 5-26555, by the DLR grants 50\,OR\,99\,036 and DLR
50\,OR\,96\,173, and by the DFG grant KO\,738/7-1. We thank Andreas Fischer
for providing us the cyclotron emission model.

\bibliography{aamnem99,/home/cthulhu/boris/tex/Papers/Bibliography/aabib}
\bibliographystyle{apj}

\section*{FIGURES}

\figcaption[]{\label{f-fuv_average_spec} 
Right: The average G140L FUV spectrum of AR\,UMa and the short G230L
NUV exposure (see Table\,\ref{t-obslog}). The mismatch between the FUV
and the NUV spectra is most likely due to the different orbital phase
sampling. Left: Blow-up of the \La\ region. Probable interstellar
absorption lines are indicated below the spectrum, absorption and
emission features intrinsic to AR\,UMa are indicated above the
spectrum.  The shaded region at the bottom gives a measure of the flux
error.}

\figcaption[]{\label{f-fuv_phase_spec}
Phase-resolved G140L spectra of AR\,UMa, orbital phases are
indicated. Note the orbital variation of the $1s_0\rightarrow2s_0$
absorption near $\sim1180$\,\AA\ and of the \La\ emission, as well as
the appearance of a broad ($\sim200$\,\AA) emission bump at
$\sim1650$\,\AA\ during $\phi=0.9-1.1$.}

\figcaption[]{\label{f-overall}
Combined ultraviolet/optical low state spectrum of AR\,UMa. The red
end of the optical spectrum is dominated by the emission of the M6
secondary star. Left: illustrative comparison of the AR\,UMa data to
non-magnetic pure hydrogen white dwarf model spectra with
$\Twd=50\,000$\,K and $\Twd=15\,000$\,K. Both models were normalized
to $V=16.9$, the (assumed) white dwarf magnitude of AR\,UMa obtained
by subtracting an appropriately scaled M6 spectrum from the observed
low state spectrum.  Right: illustrative comparison of the AR\,UMa
data with $\Twd=25\,000, 20\,000, 15\,000$\,K (top to bottom) magnetic
white dwarf model spectra ($B=200$\,MG), again normalized to
$V=16.9$.}

\figcaption[]{\label{f-fuv_mwd} Dashed line: magnetic white dwarf model
spectrum with $\Twd=17\,000$\,K, $B=200$\,MG and
$\Rwd=8.2\times10^8$\,cm (at $d=88$\,pc). Full line: {\it STIS\/} FUV
data of AR\,UMa at $\phi=0.5$ with the $1445$\,\AA\ cyclotron
component (Sect.\,\ref{s-cyc_model}) subtracted. The
$1s_0\rightarrow2s_0$ transition is discussed in detail in
Sect.\,\ref{s-1180}.}

\figcaption[]{\label{f-cyc_model}
Cyclotron emission in AR\,UMa. Left panel: FUV spectrum for $\phi=0.5$
(topmost curve) and the cyclotron model derived from the 1445\,\AA\
emission bump (bottom curve). The difference of the observed FUV
spectrum and the cyclotron model is shifted downwards by 0.5 units.
The only available NUV spectrum is also shown, but does not match in
phase ($\phi=0.08$). Right panel: At $\phi\sim0.0$ the FUV spectrum
contains an additional emission bump centered at $\sim1650$\,\AA,
which we attribute to cyclotron emission from the southern accretion
region. Plotted are the observed FUV/NUV spectra centered on/near
$\phi\sim0.0$ (topmost curves), the cyclotron model for the 1650\,\AA\
bump (bottom curve), and the difference of the FUV ($\phi\sim0.0$)
spectrum and {\it both\/} cyclotron models (shifted downwards by 0.5
units).}

\figcaption[]{\label{f-lyalpha}
Variable \La\ absorption/emission. Top panel: the observed FUV
spectrum at $\phi=0.9$ (solid line) and a damping profile
corresponding to an interstellar column density
$\Nh=1.1\times10^{18}\mathrm{cm^{-2}}$. Bottom: enlargement of the
phase-resolved spectra around \La, corrected for the interstellar
absorption as displayed above. The individual spectra (from bottom to
top) have been shifted upwards by 0, 3, 14, 21, and 23 units.}

\figcaption[]{\label{f-lyalpha_flux}
Velocity (top) and flux (bottom) of the \La\ emission line. The top
panel shows also the best-fit sine curve to the velocity variation.}

\figcaption[]{\label{f-lightcurves}
FUV light curves extracted from the time-tagged {\it STIS\/} photon
stream in four wavelength bands. From top to bottom:
1146.5--1206.3\,\AA, 1223.9--1296.9\,\AA, 1323.2--1520.3\,\AA,
1529.1--1717.5\,\AA. Shown as solid lines are sine fits to the data
with the parameters listed in Table\,\ref{t-lc_par}.}

\figcaption[]{\label{f-modulation_amplitude}
Top: relative modulation of the FUV emission of AR\,UMa in
$\sim30$\,\AA\ bands. Bottom: mean FUV spectrum of AR\,UMa.}

\figcaption[]{\label{f-power}
Power spectra of the G140L data of AR\,UMa (top curve) and of a
simulated sine wave with a mean count rate, amplitude, period, and
coverage as defined by the STIS data (bottom curve, shifted down by 4
units).}

\begin{figure*}
\includegraphics[angle=270,width=18cm]{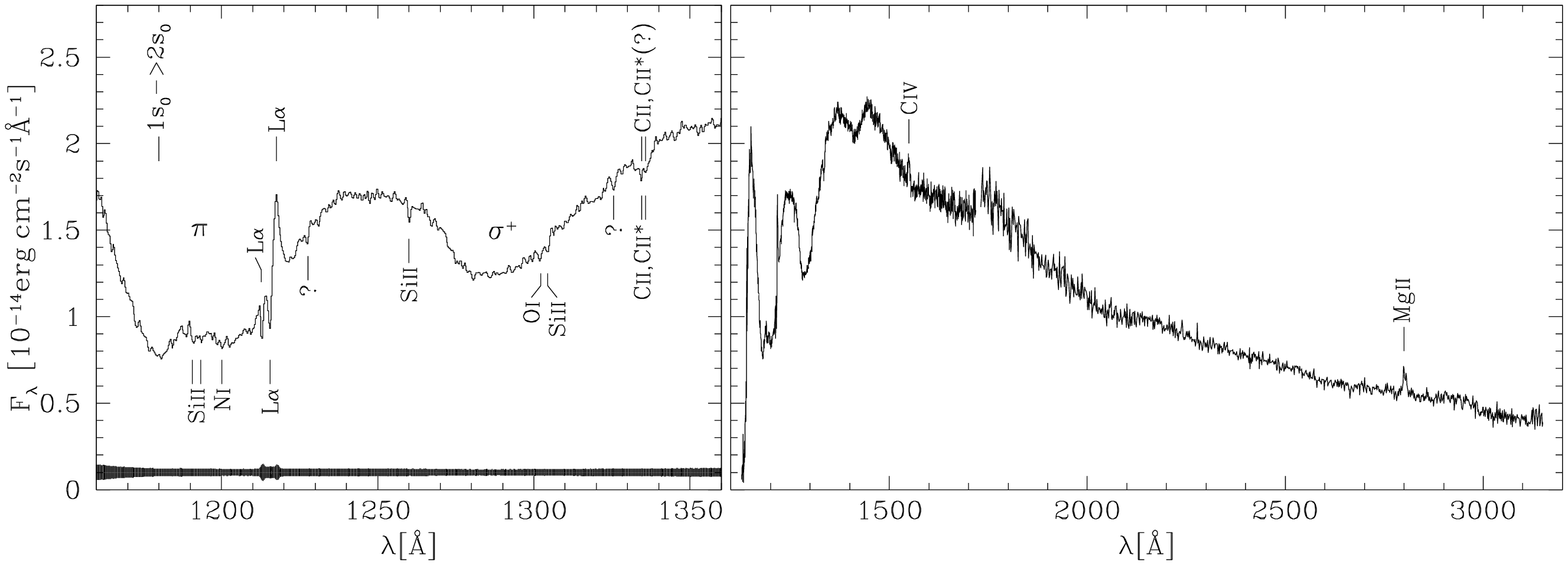}
\centerline{Figure \ref{f-fuv_average_spec}}
\end{figure*}

\begin{figure}
\includegraphics[width=8.8cm]{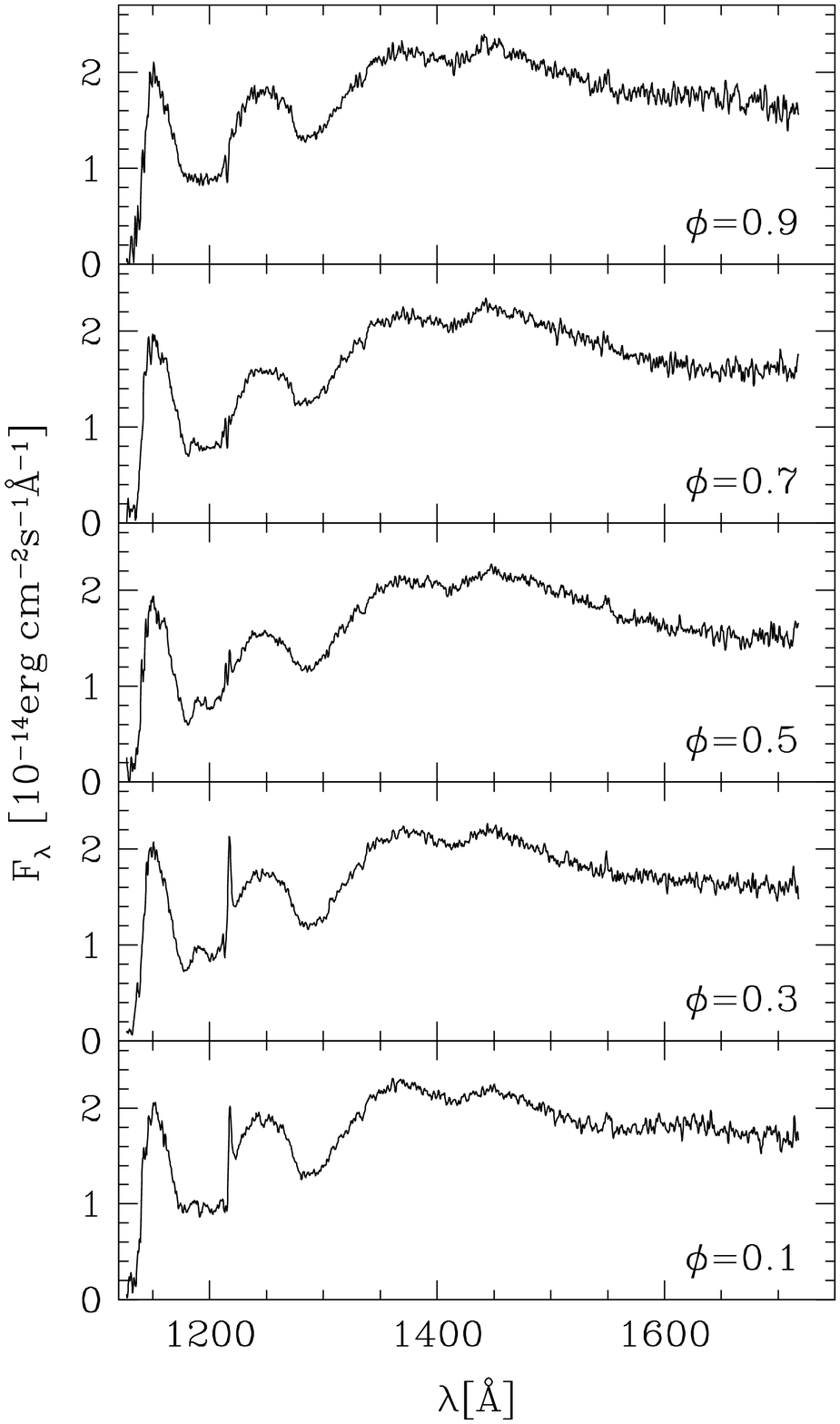}\\
\parbox{8.8cm}{\centerline{Figure \ref{f-fuv_phase_spec}}}
\end{figure}

\begin{figure*}
\includegraphics[angle=270,width=8.8cm]{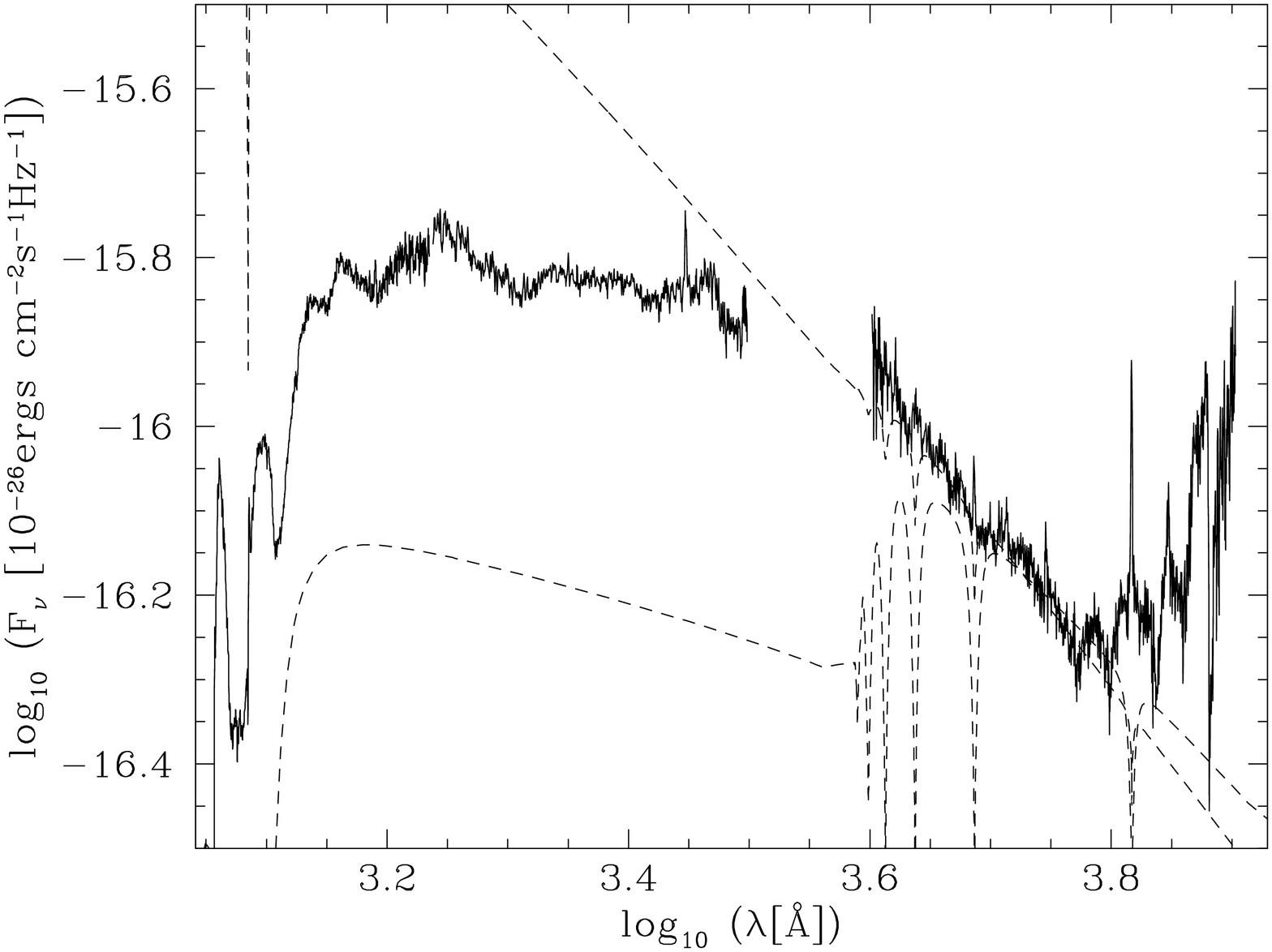}
\includegraphics[angle=270,width=8.8cm]{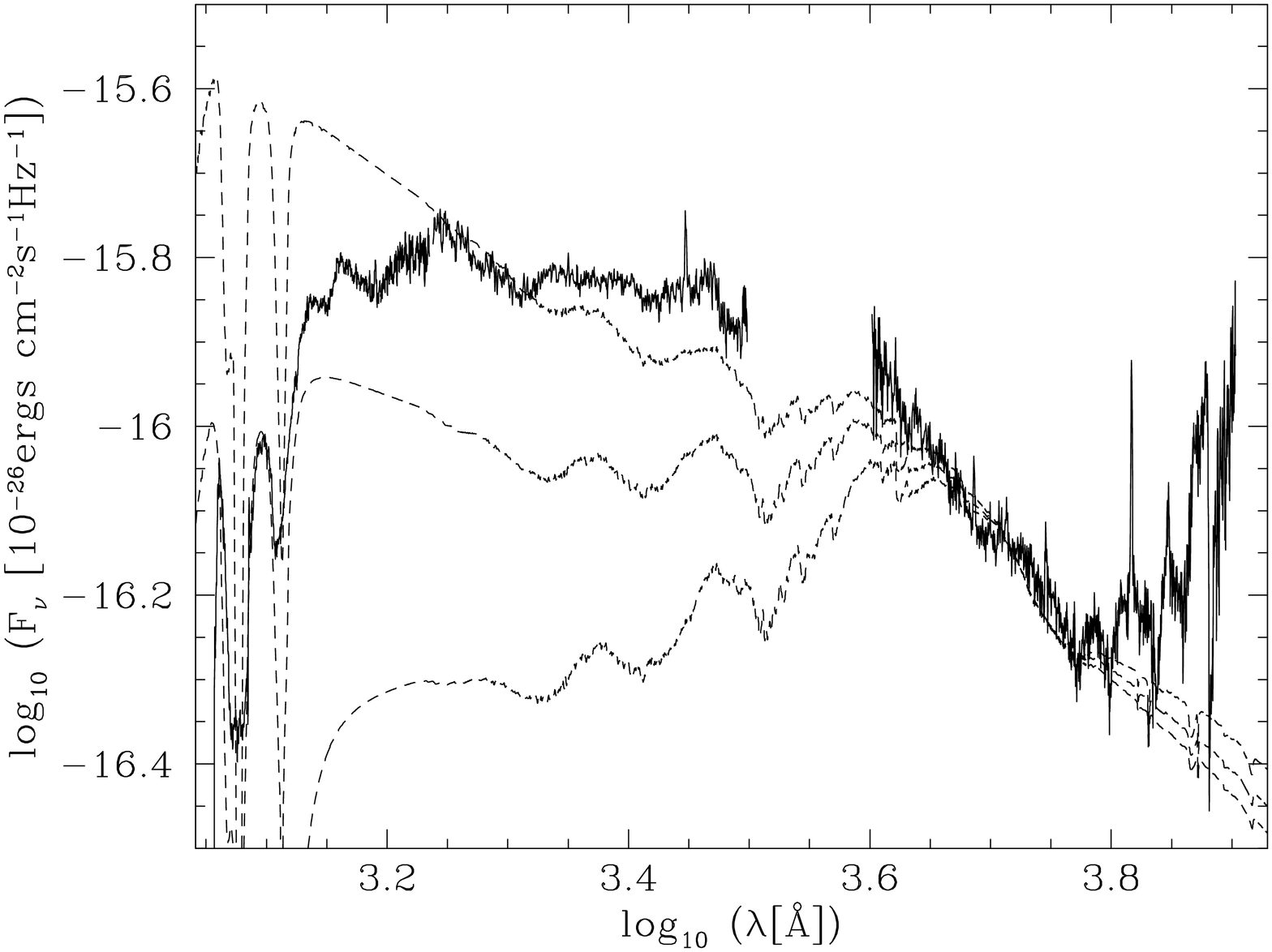}
\centerline{Figure \ref{f-overall}}
\end{figure*}

\begin{figure}
\includegraphics[angle=270,width=8.8cm]{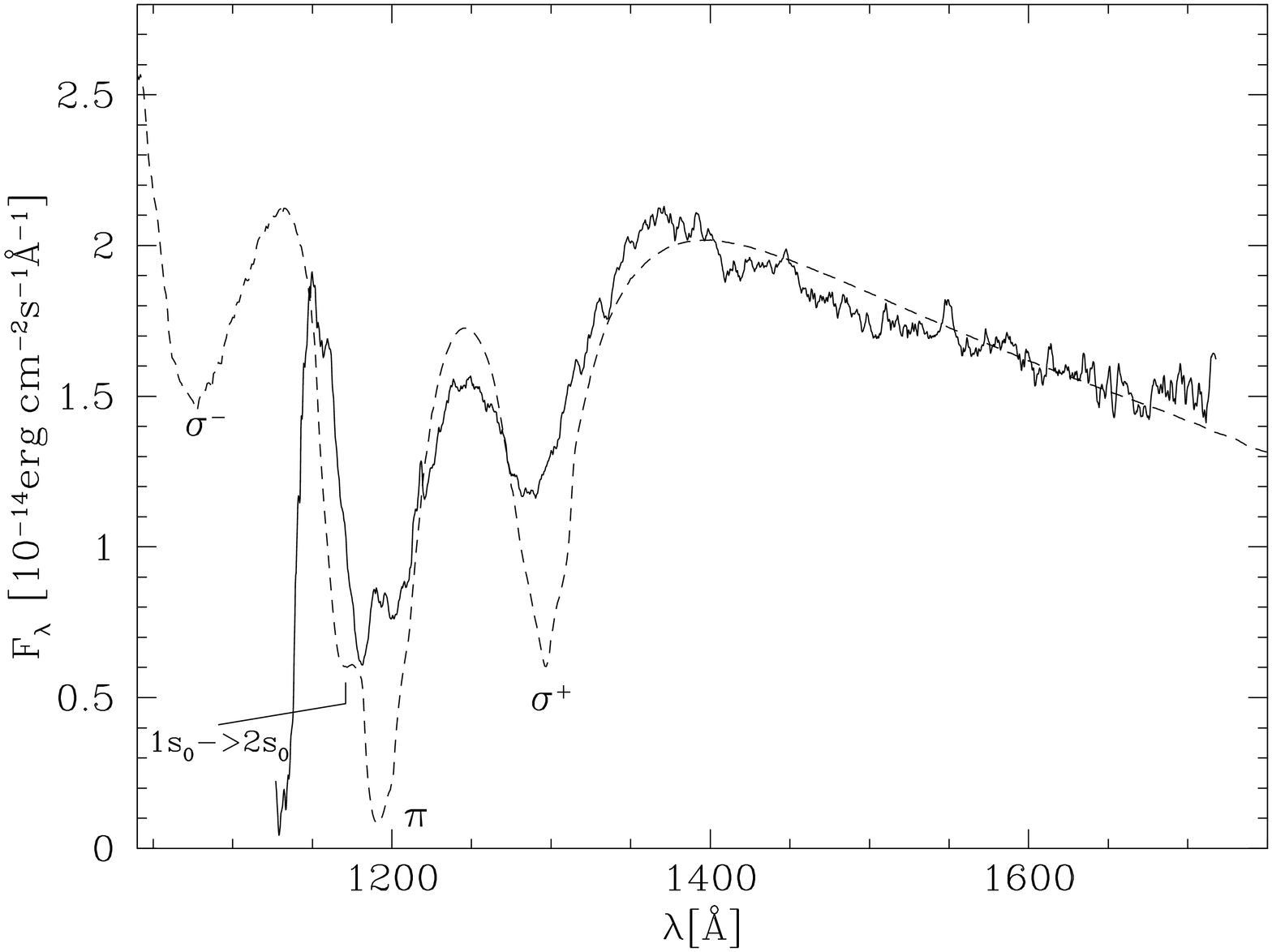}\\
\parbox{8.8cm}{\centerline{Figure \ref{f-fuv_mwd}}}
\end{figure}

\begin{figure*}
\includegraphics[angle=270,width=8.8cm]{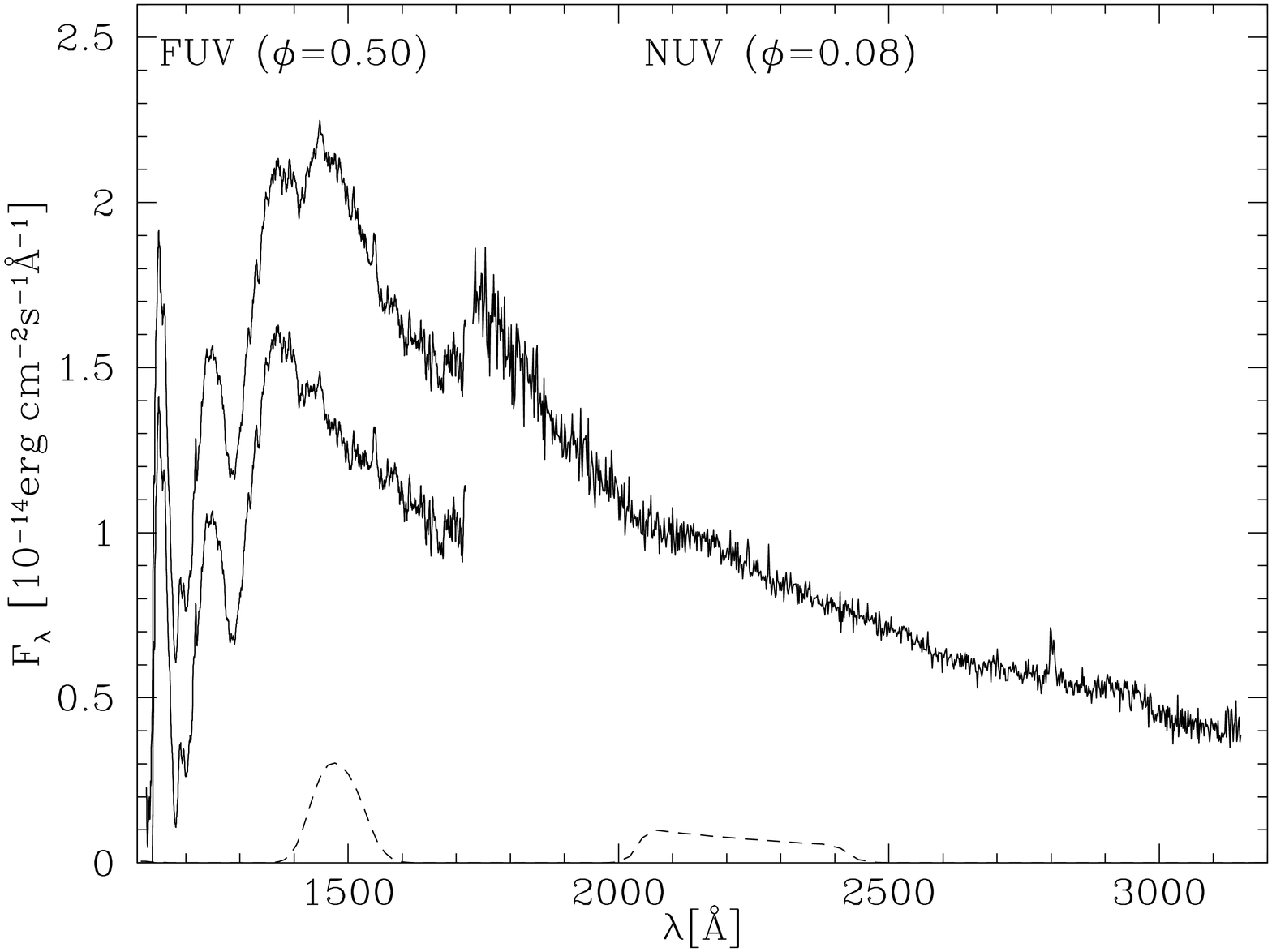}
\includegraphics[angle=270,width=8.8cm]{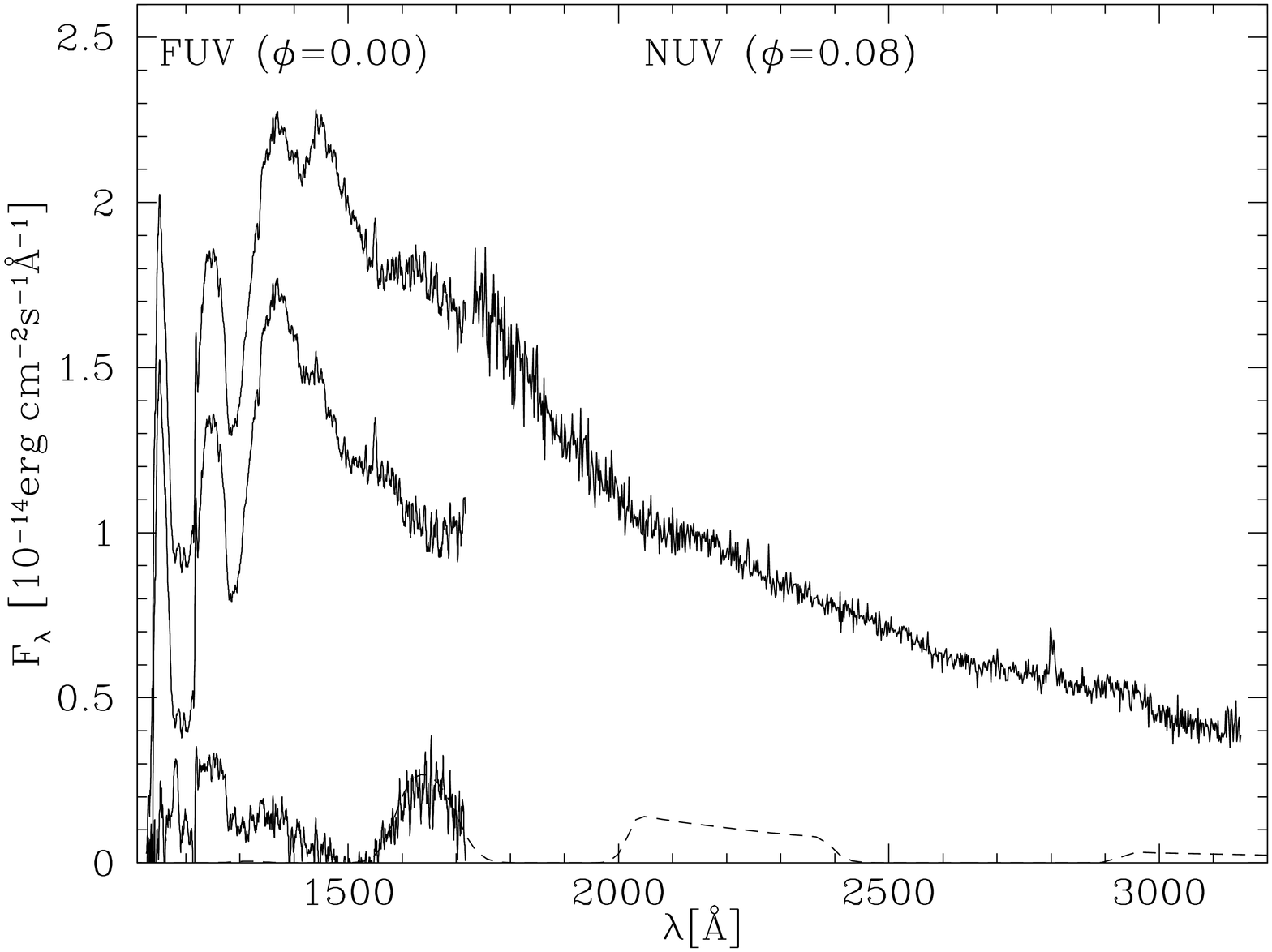}
\centerline{Figure \ref{f-cyc_model}}
\end{figure*}

\begin{figure}
\includegraphics[width=8.8cm]{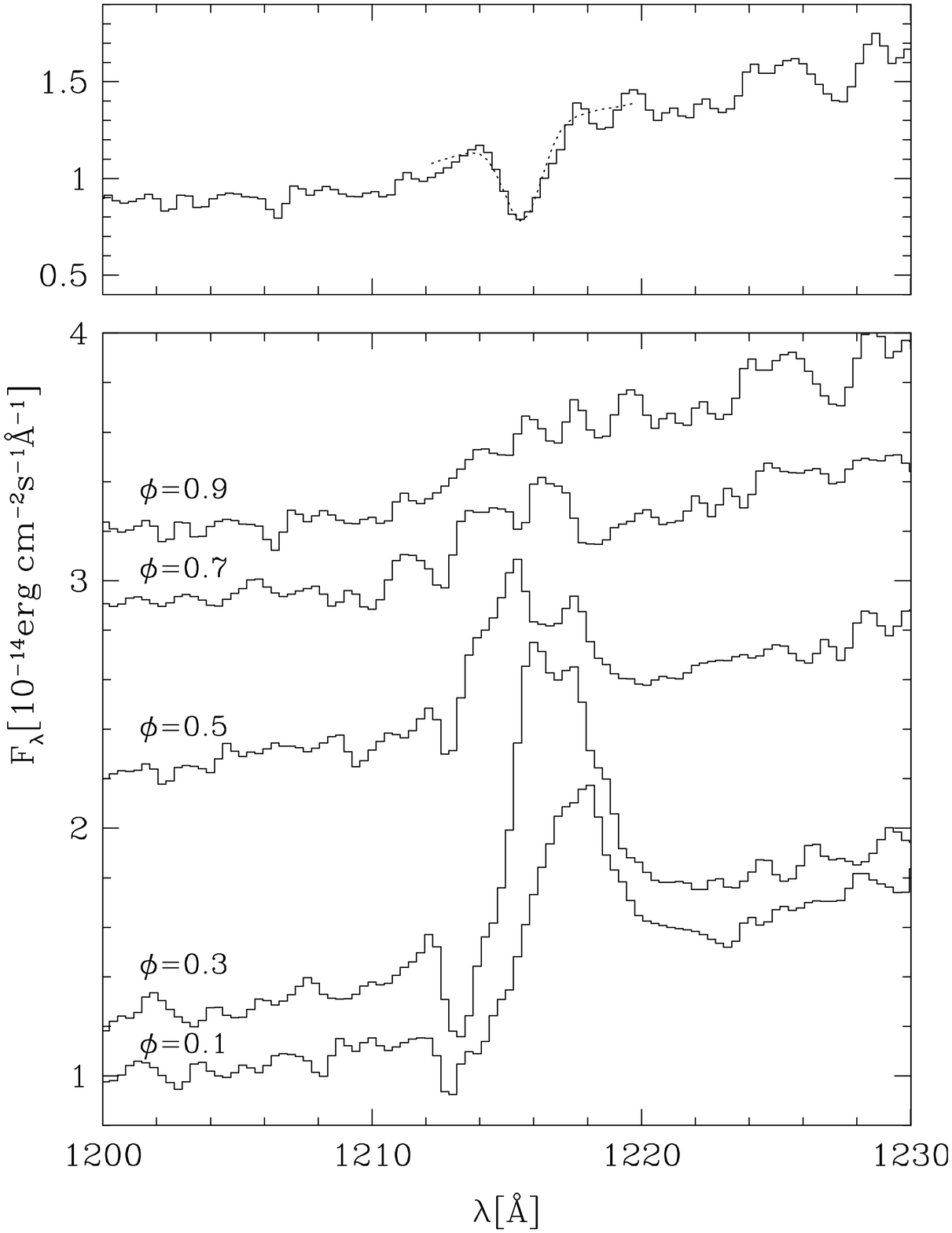}\\
\parbox{8.8cm}{\centerline{Figure \ref{f-lyalpha}}}
\end{figure}

\begin{figure}
\includegraphics[angle=270,width=8.8cm]{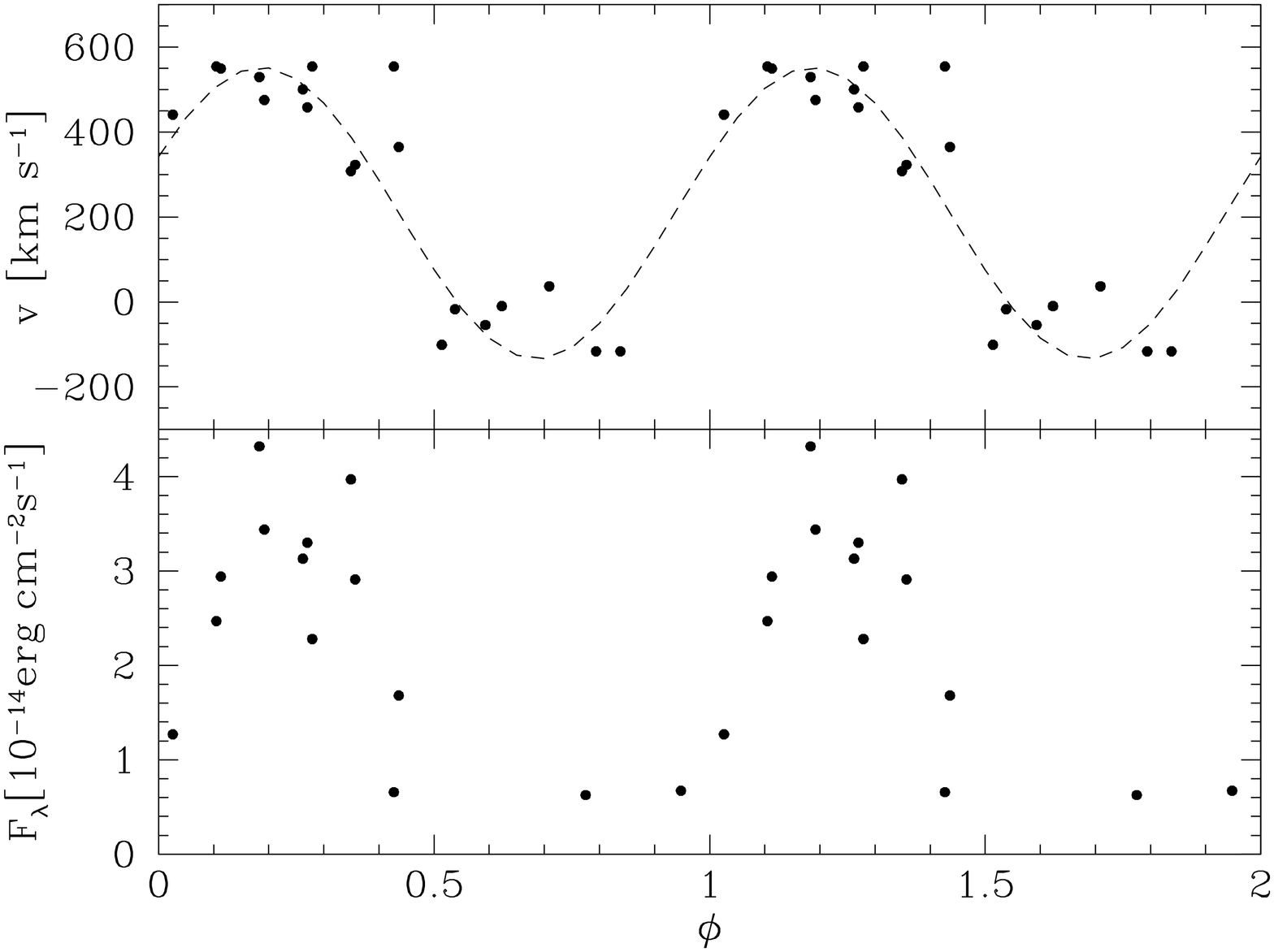}\\
\parbox{8.8cm}{\centerline{Figure \ref{f-lyalpha_flux}}}
\end{figure}

\begin{figure}
\includegraphics[width=8.8cm]{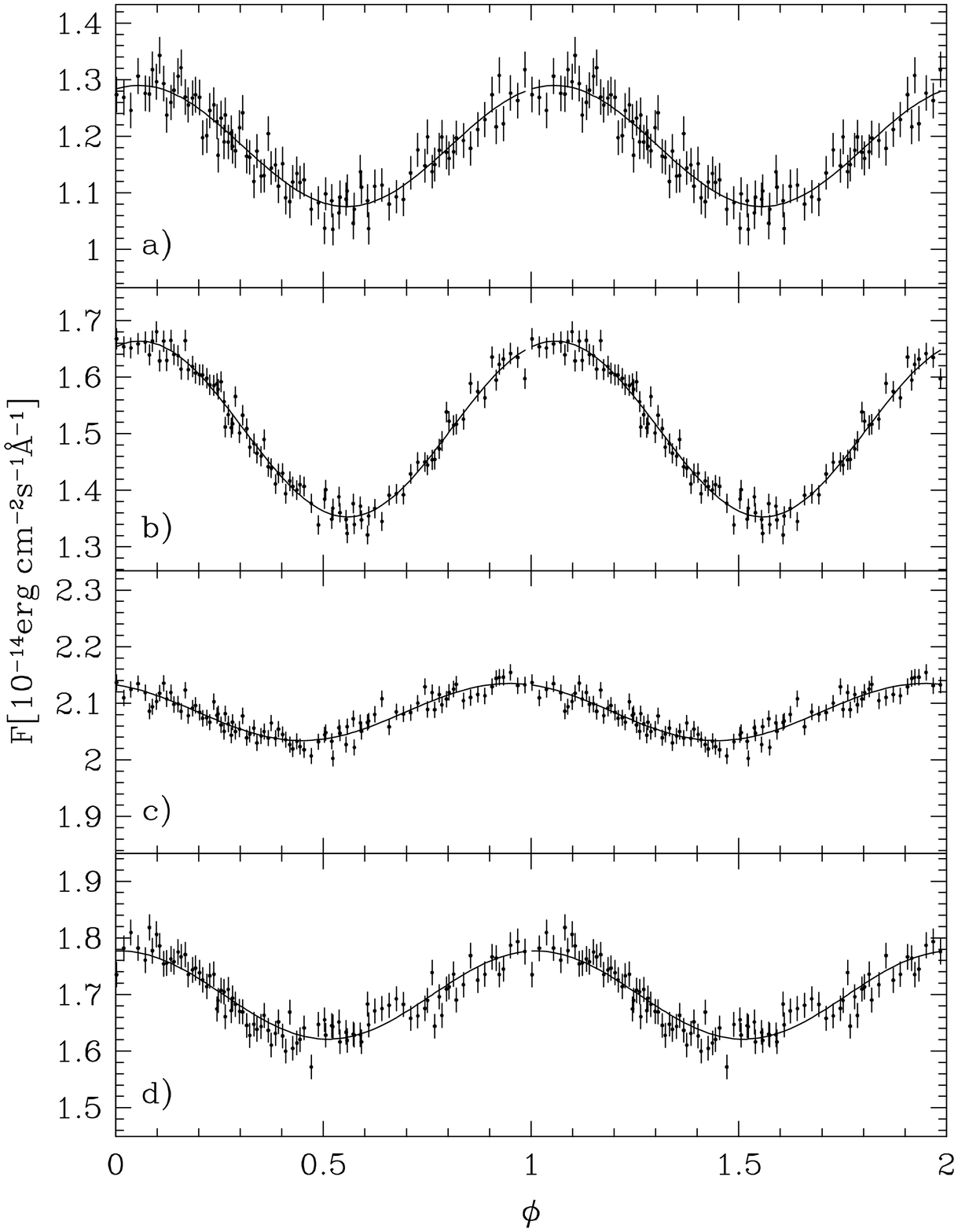}\\
\parbox{8.8cm}{\centerline{Figure \ref{f-lightcurves}}}
\end{figure}

\begin{figure}
\includegraphics[angle=270,width=8.8cm]{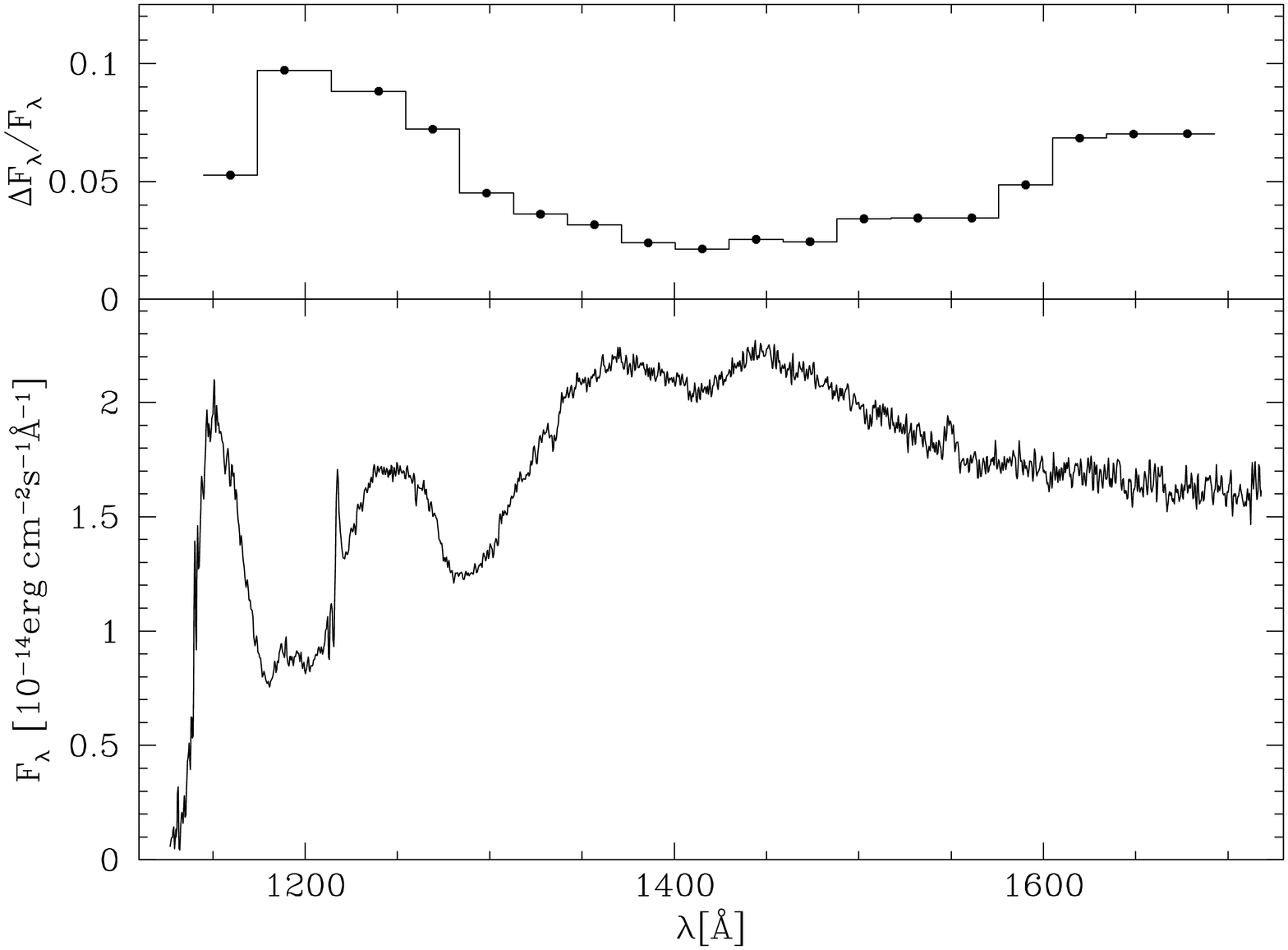}\\
\parbox{8.8cm}{\centerline{Figure \ref{f-modulation_amplitude}}}
\end{figure}

\begin{figure}
\includegraphics[angle=270,width=8.8cm]{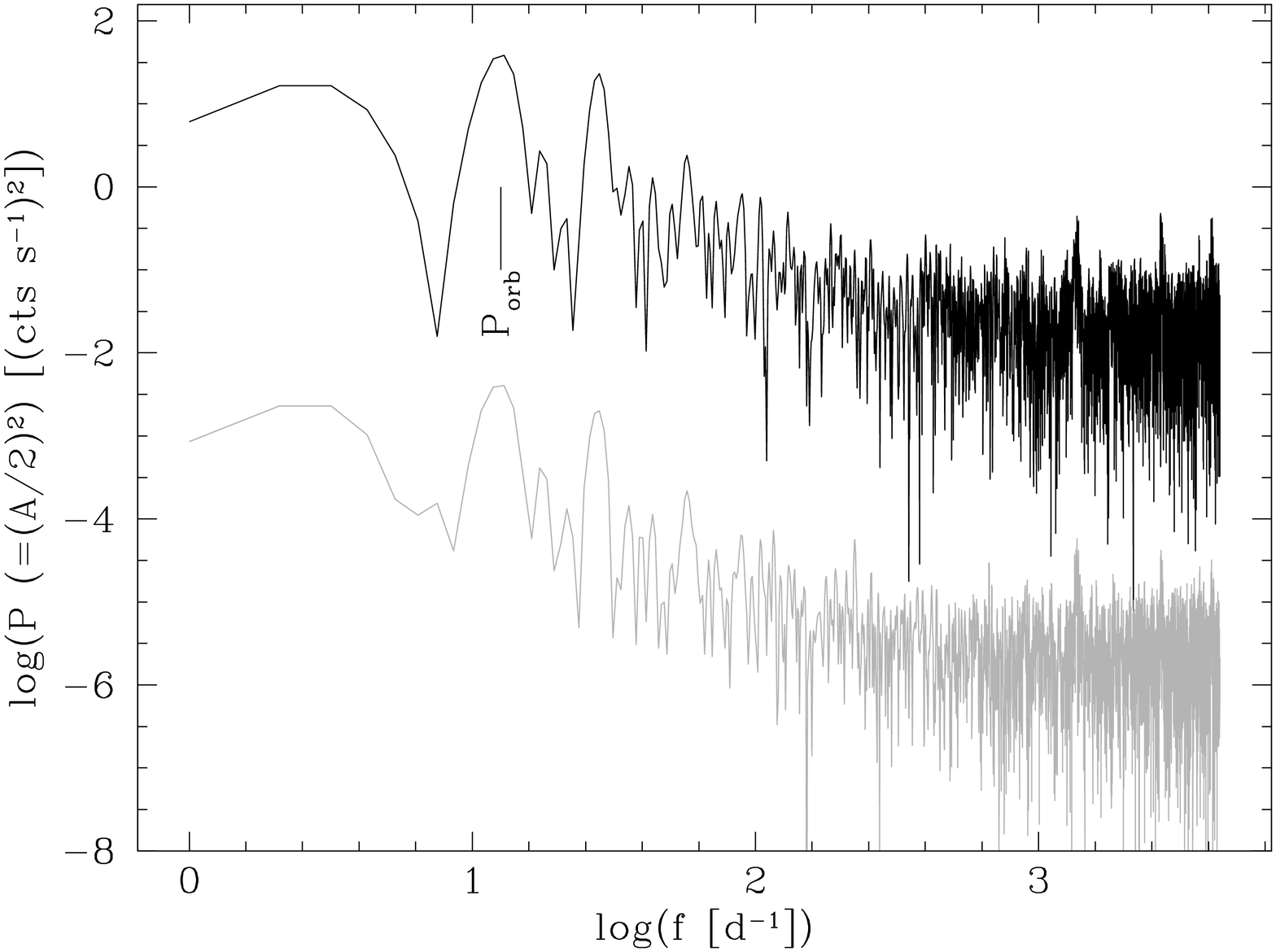}\\
\parbox{8.8cm}{\centerline{Figure \ref{f-power}}}
\end{figure}

\newpage

\begin{deluxetable}{rrrrrrrr}
\tablecolumns{5}  
\tablewidth{0pc}  
\tablecaption{\label{t-obslog}Log of the {\it HST\/} observations on 1998 December 9}
\tablehead{  
\colhead{Dataset} &
\colhead{Grating} &
\colhead{Obs. start} &
\colhead{Exp. time} &
\colhead{$\Delta\phi$}\\
\colhead{} &
\colhead{} &
\colhead{(UT)} &
\colhead{sec} &
\colhead{}}
\startdata  
o53y01010 &G140L& 13:26:41 & 2370 & 0.499 -- 0.826\\ 
o53y01020 &G140L& 14:52:54 & 2730 & 0.242 -- 0.622\\ 
o53y01030 &G140L& 16:29:39 & 2730 & 0.077 -- 0.457\\
o53y01040 &G140L& 18:06:25 & 2730 & 0.912 -- 0.291\\
o53y01050 &G140L& 19:43:11 & 1320 & 0.747 -- 0.928\\
o53y01060 &G230L& 20:13:24 & 1200 & 0.987 -- 0.159\\
\enddata  
\end{deluxetable}  

\begin{deluxetable}{rrrrrrrr}  
\tablecolumns{5}  
\tablewidth{0pc}  
\tablecaption{\label{t-lc_par}
Wavelength bands of the ultraviolet light curves and
best-fit sine parameters.}
\tablehead{  
\colhead{Band} &
\colhead{$\lambda$} & 
\colhead{$A$} & 
\colhead{$O$} & 
\colhead{$\phi_\mathrm{max}$}\\
 & 
\colhead{[\AA]} &
\multicolumn{2}{c}{$[10^{-14}\mathrm{erg\,cm^{-2}s^{-1}\AA^{-1}}]$}}
\startdata
a & 1146.5 - 1206.3 & 0.107 & 1.183 & 0.806 \\ 
b & 1223.9 - 1296.9 & 0.155 & 1.508 & 0.808 \\ 
c & 1323.2 - 1520.3 & 0.050 & 2.085 & 0.698  \\
d & 1529.1 - 1717.5 & 0.078 & 1.699 & 0.761  \\
\enddata  
\end{deluxetable}

\end{document}